\documentclass{ws-procs975x65}

\newcommand{\BEQ}{\begin{eqnarray}}
\newcommand{\EEQ}{\end{eqnarray}}
\newcommand{\BEA}{\begin{eqnarray}}
\newcommand{\EEA}{\end{eqnarray}}

\begin{document}

\title{GRAVITATIONAL HYDRODYNAMICS VS OBSERVATIONS OF VOIDS, JEANS CLUSTERS
AND MACHO DARK MATTER}


\author{Theo M. Nieuwenhuizen$^{1}$, Carl H. Gibson$^{2}$,  and Rudolph E. Schild$^{3}$}

\address{$^1$ Institute for Theoretical Physics, 
Valckenierstraat 65, 1018 XE Amsterdam, the Netherlands\\
$^2$  Mech. and Aerosp. Eng. \& Scripps Inst. Oceanogr. Depts., UCSD, 
 La Jolla, CA 92093, USA\\
$^3$ Harvard-Smithsonian Center for Astrophysics,
       60 Garden Street, Cambridge, MA 02138, USA}

\begin{abstract}
Gravitational hydrodynamics acknowledges that hydrodynamics is essentially nonlinear and viscous.
In the plasma, at $z=5100$, the viscous  length enters the horizon and causes fragmentation into plasma clumps
surrounded by voids. The latter have expanded to $38$ Mpc now, explaining the cosmic
void scale $30/h=42$ Mpc. After the decoupling the Jeans mechanism fragments all matter in clumps
of ca 40,000 solar masses. Each of them fragments due to viscosity in millibrown dwarfs
of earth weight, so each Jeans cluster contains billions of them. The Jeans clusters act as
ideal gas particles in the isothermal model, explaining the flattening of rotation curves.
The first stars in old globular clusters are formed by aggregation of milli brown dwarfs, without dark period.
Star formation also happens when Jean clusters come close to each other and agitate and heat up the
cooled milli brown dwarfs, which then expand and  coalesce to form new stars.
This explains the Tully-Fischer and Jackson-Faber relations, and the formation of young globular
clusters in galaxy mergers. Thousand of milli brown dwarfs have been observed in quasar microlensing
and some 40,000 in the Helix planetary nebula.

While the milli brown dwarfs, i.e., dark baryons, constitute the galactic dark matter,
cluster dark matter consists probably of 1.5 eV neutrinos, free streaming at the decoupling.
These two types of dark matter explain a wealth of observations.
\end{abstract}

\bodymatter

\section{Introduction}
It is generally understood that hydrodynamics is in principle important in the early Universe,
but its role is small in practice.
There is a great impetus from quantum field theory, inflation of a mysterious scalar field
explains the structure in the CMB. In this approach, hydrodynamics is linearized before the
decoupling of photons and matter.
It was Gibson 1996 who questioned this assumption~\cite{Gibson96}. The conclusion is astonishing:
{\it There is a viscous instability in the plasma, so the concordance approach cannot be correct.
Hydrodynamics can explain structure formation without cold dark matter trigger.}
Here we summarize our recent work,  Ref.~\citen{NGS09}.

\section{Hydrodynamics}

The Navier-Stokes eqn expresses conservation of the specific momentum in a fluid,

\begin{equation}
\label{NavStokes}
\frac{\partial \vec v }{ \partial t} =
\nabla B + \vec v\! \times \vec \omega
+ \vec F_{viscous} +  \vec F_{other},
\end{equation}
averaged over system control volumes exceeding the momentum
collision length scale. $B = p/\rho +  \frac{1}{2} v^2 + lw$
is the Bernoulli group of mechanical energy terms.
For adiabatic flows the ``lost work'' term $lw$ due to frictional
losses is negligible so the enthalpy $p/\rho$ decreases or
increases to compensate for changes in the kinetic energy per unit
mass $ \frac{1}{2} v^2$.  The viscous force
is $\vec{F}_{viscous}=\nu_s\nabla^2\vec{v}+(\frac{1}{3}\nu_s
+\nu_b)\nabla\cdot(\nabla\cdot\vec{v})$, with kinematic shear
viscosity $\nu_s=\eta/\rho$ and bulk viscosity $\nu_b=\zeta/\rho$,
while other fluid forces may arise. The inertial-vortex force per
unit mass $ \vec v \times \vec \omega$, with
$\vec\omega=\nabla\!\times\vec v$, produces turbulence if it
dominates the other forces; for example, $|\vec v
\times \vec \omega| / |\vec F_{viscous}|$ is the Reynolds number.
A large viscosity corresponds to a small Reynolds number, with
universal critical value  $\sim 25-100$.

Jeans considers a gas clump of density $\rho_B$, typical size $L$
and mass $\rho_B L^3$. Its gravitational force has magnitude $ G(\rho_B L^3)^2/L^2$.
Equating this to the $\vec v \times \vec \omega $ force
$(\rho_B L^3)v^2/L$ with $v$ the sound speed, yields the Jeans length $L_J=v/\sqrt{G\rho_B}$.
A fluid will collapse on scales of $L_J$ into mass clumps $\rho_B L_J^3$.
It is generally believed that only this scale is relevant, but Gibson considers the case
where viscosity yields the dominant force~\cite{Gibson96}.
Applying the balance of gravitation and viscous force $\sim (\rho_B L^3)\nu v/L^2$ brings
the Schwarz viscous scale $L_{SV}=(\nu v/G\rho_B)^{1/3}$. It appears that merely
these two scales suffice to explain the major properties of cosmic structure formation.

 We consider a flat cosmology and  $h=0.744$~\cite{Nneutrino09},
 so that neutrino dark matter has
$\Omega_\nu=0.111/$$h^{3/2}$$=0.173$, while no cold dark matter is assumed to exist.
For baryons we take $\Omega_B=0.02265/h^2=0.0409$ (WMAP5) and
 $\Omega_\Lambda=0.786$ assures a flat space.

\section{Viscous instability in the plasma}

Silk damping involves a photon mean free path of $10^{-5}$ pc at decoupling, while
the acoustic horizon is $d_H^{ac}=128$ kpc, so it is no surprise that the viscous lenght
$L_{SV}=76$ kpc is smaller already.
Indeed, the shear viscosity increases in time as $1/T^2$,  and before
decoupling the plasma becomes too viscous to follow the cosmic expansion~\cite{Gibson96,NGS09}.
A viscous instability occurs at $z=5120$ when $L_{SV}$ enters the horizon $d_H^{ac}$.
It tears the plasma apart at density minima, and creates voids, next to condensations
with baryonic clustering mass ${\pi}\rho_BL^3/6=1.7 \cdot 10^{14}M_\odot$, fat galaxy clusters. 
The initial void scale $d_H^{ac}=7.3$ kpc expands by a factor
$1+z$ to become 38 Mpc now, a typical void size, $30/h$ Mpc = 40 Mpc.
$\Lambda$CDM predicts such voids formed last and
full of debris, rather than first and empty as observed.
At decoupling the neutrino hot dark matter is still free streaming and homogeneous.
At  $z\sim7$ the neutrinos condense on e.g. galaxy clusters
so the baryonic voids, still filled with neutrinos, become completely empty. ~\cite{Nneutrino09}
Thus till $z\sim7$ the metric is quite uniform.

\section{Fragmentations in the gas at the Jeans and viscous scales}

At last scattering, the plasma turns into a neutral gas and further
baryonic structures form.  For H with 24\% weight
of He, the sound speed is $v=5.68$ km/s. The gas
fragments at the Jeans scale into clumps of mass $M_{J}=\pi\rho_{B}L_J^3/6=
38,000\,M_\odot$.

At decoupling the viscosity decreases
to the $10^{13}$ times smaller hot-gas value. The viscous scale
$L_{SV}=0.14$ pc  m implies a
further condensation of Jeans clumps into masses of order
$\pi\rho_B L_{SV}^3/6=4\cdot10^{-5}M_\odot=13 M_\oplus$.
We call these objects milli brown dwarfs (mBDs),
each Jeans cluster contains billions of them.

\section{Comparison to observations}

In some of the Jeans clusters (JCs) collision processes quickly lead
to star formation, basically without a dark period, thus
transforming them into globular clusters and ordinary
stars. In the major part of the JCs the  mBDs cool and
they still persist without stars. These JCs are in mainly in isothermal equilibrium
and act as ideal gas particles that constitute the galactic dark
matter. Their physical presence explains why the isothermal model
describes flattening of the rotation curves, while the Tully-Fischer and
Faber-Jackson relations follow from JC-JC interactions~\cite{NGS09}.

The matrix of dark  JCs  is revealed by new
star formation (young globular clusters) as seen in photographs of
galaxy mergers such as Tadpole, Mice and Antennae.

{}From the GHD scenario following decoupling, the first stars form
gently by a frictional binary accretion of still warm  mBDs to form
larger planet pairs and finally small stars as observed in old globular clusters.
Thereby they create an Oort cavity as clearly exposed in e. g.
the Helix planetary nebula.

At decoupling the entire baryonic universe thus turns to a fog of mBDs.~\cite{Gibson96}
Quasar microlensing observations~\cite{Schild96} confirm that galactic dark matter
is composed from such objects. Indeed, thousands of crossings events
have been observed, of mostly earth mass objects and some Jupiters.
Due to the cosmic expansion the mBDs cool and the
freezing temperature of H and He occurs at redshift $z
\approx 8$, producing the cool dark baryonic  mBDs  in clumps predicted as the galaxy
dark matter. ~\cite{Gibson96}

Cluster dark matter is the ``true'' dark matter, probably composed of 1.5 eV
neutrinos. They condense at $z\sim 7$ on galaxy clusters, causing the extra-galactic
  mBDs  to reionize into  hot X-ray gas~\cite{Nneutrino09,TheoMG12}.
Since then the voids are fully empty and the universe is strongly inhomogeneous.

When cooled, the   mBDs   are too small to dim light, even
from remote sources, but they can account for dimming when they
are heated. Warm atmosphere diameters are $\approx 10^{13}$ m, the
size of the solar system out to Pluto, bringing them out of the
dark. The separation distance between  mBDs  is $\approx 10^{14}$
m, as expected if the  JC  density of  mBDs  is the primordial
density $\rho_0 = 2 \times 10^{-17}$ kg m$^{-3}$. In planetary
nebula such as the nearby Helix, dark  mBDs  at the
boundary of the $3 \times 10^{15}$ m Oort cavity are evaporated.
HST optical images of Helix show $\approx 10^4$  ``cometary
knots'', gaseous planet-atmospheres $\approx 10^{13}$ m  which we identify
as   mBDs   with metallicity, and Spitzer shows $40,000$ of them in
the infrared from the $10^3 M_{\odot}$ available.
Similar behavior is observed in other planetaries.

The large density contrast between plasma clumps and voids does not
prevent a $10^{-5}$ CMB temperature contrast, since
temperature is mainly set by the expansion.~\cite{NGS09}

\end{document}